\documentclass[conference]{IEEEtran}
\usepackage{graphicx}
\usepackage{amsfonts}
\usepackage{amsmath}
\usepackage{float}
\usepackage{multirow}
\usepackage{color}
\usepackage{enumitem}
\usepackage{graphicx}
\usepackage{url}
\usepackage{algorithmic}
\usepackage[switch, columnwise]{lineno}
\usepackage{epstopdf}
\usepackage{balance}
\usepackage[ruled,vlined]{algorithm2e}
\usepackage{amsmath}
\usepackage{amsthm}

\usepackage{colortbl}
\usepackage{xcolor}
\usepackage{array}
\usepackage{graphicx}
\usepackage{booktabs}
\usepackage{subcaption}

\usepackage{booktabs}
\usepackage{amsthm}
\newtheorem{theorem}{Theorem}[section]
\newtheorem{corollary}{Corollary}[theorem]

\newtheorem{definition}{Definition}[section]

\begin{document}

\title{Achieving Unanimous Consensus Through Multi-Agent Deliberation}

\author{
    \IEEEauthorblockN{\textsuperscript{}  Apurba Pokharel}
    \IEEEauthorblockA{
    \small{University of North Texas}\\
    apurba.pokharel@unt.edu}
    \and
    \IEEEauthorblockN{\textsuperscript{} Ram Dantu}
    \IEEEauthorblockA{
    \small{University of North Texas}\\
    ram.dantu@unt.edu}
    
    \and
    \IEEEauthorblockN{\textsuperscript{} Shakila Zaman}
    \IEEEauthorblockA{
    \small{University of North Texas}\\
    shakila.zaman@unt.edu}
    \and
    \IEEEauthorblockN{\textsuperscript{}  Vinh Quach}
    \IEEEauthorblockA{
    \small{University of North Texas}\\
    vinh.quach@unt.edu}
    \and
    \IEEEauthorblockN{\textsuperscript{} Sirisha Talapuru}
    \IEEEauthorblockA{
    \small{University of North Texas}\\
    sirishamadh.talapuru@unt.edu}
}

\maketitle

\begin{abstract}
Blockchain consensus mechanisms have relied on algorithms such as Proof-of-Work (PoW) and Proof-of-Stake (PoS) to ensure network functionality and integrity. However, these approaches struggle with adaptability for decision-making where the opinions of each matter rather than reaching an agreement based on honest majority or weighted consensus. This paper introduces a novel deliberation-based consensus mechanism where Large Language Models (LLMs) act as rational agents engaging in structured discussions to reach a unanimous consensus. By leveraging graded consensus and a multi-round deliberation process, our approach ensures unanimous consensus for definitive problems and graded consensus for prioritized decision problems and policies. We provide a formalization of our system and use it to show that the properties of blockchains are maintained, while also addressing the behavior in terms of adversaries, stalled deliberations, and confidence in consensus. Moreover, experimental results demonstrate system feasibility, showcasing convergence, block properties, and accuracy, which enable deliberative decision-making on blockchain networks. 
\end{abstract}

\section{\textbf{Introduction and Motivation}}
\begin{figure}[htp!]
    \includegraphics[width=\linewidth]{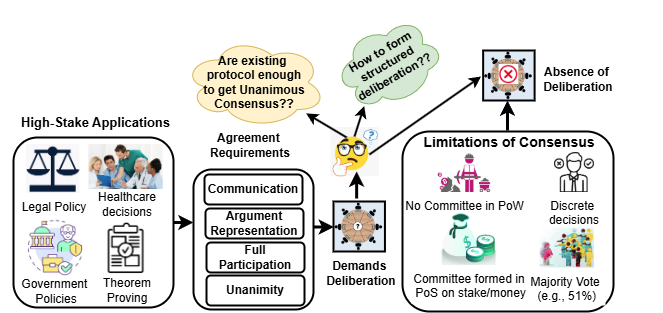}
    \caption{Decision-making process of high-stake application requires active communication, argument processing to reach a unanimous outcome, and demands a structured deliberation framework. However, existing blockchain consensus protocols are limited in addressing such complex consensus problems and fail to support deliberation among participants. }
    \label{fig:Figure 1}
    \vspace{-1em}
    \end{figure}

\vspace{-.5em}
\textbf{Absence of Deliberative Consensus in Blockchain:}
Blockchains introduced a new way of processing financial transactions within a decentralized network, which guarantees that transactions are validated and recorded by the network through a consensus mechanism (e.g.,  PoW \cite{Bitcoin}, PoS \cite{Polkadot}, etc. ). However, a key observation is made that the majority of the consensus protocols being used in blockchains are designed for cryptocurrencies \cite{ConsensusReviewPaper1} or financial applications \cite{zaman2023seamless}, \cite{ConsensusReviewPaper2}, \cite{zaman2024indivisible}. Moreover, these protocols are primarily designed with an agreement quorum to handle the presence of faulty/malicious nodes. For instance, PoW considers a 51\% majority, whereas PoS-BFT practices 33\% adversary tolerance \cite{Laport}. Although these kinds of agreements are advantageous when considering failure or malice, they cannot guarantee a reliable solution for high-stakes applications (e.g., court verdict systems, healthcare decisions, ethics committees, and government policy-making systems) that require unanimous consensus among all participants  \cite{pokharel2024deliberation}. For instance, a jury should reflect a fair verdict by evaluating all committee members' opinions through rounds of structured deliberation to prevent biased or unfair decisions \cite{salerno2010promise}. In healthcare, decisions such as transitioning to human drug testing require strong or near-unanimous agreement, which is likewise impossible with current consensus mechanisms. Figure \ref{fig:Figure 1} represents the limitations of current consensus protocols.

\begin{figure*}[t]
    \centering
    \includegraphics[width=\textwidth]{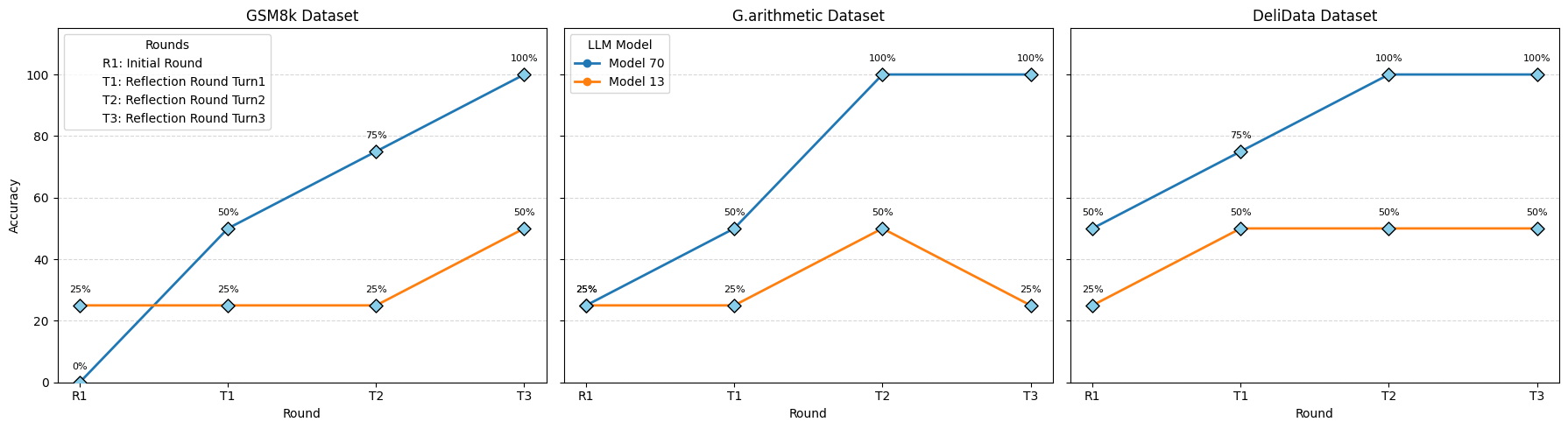}
    \caption{Consensus accuracy of different LLMs through rounds of deliberation. Large parameter models (e.g., 70b) benefit more from reflection and reinforce their reasoning as seen by the linear increase in accuracy. Whereas smaller models (e.g., 13b) are unable to distinguish between right and wrong answers, and thus, accuracy plateaus or even decreases. This mirrors real-world deliberation, where subject matter experts are more likely to reach consensus and improve collective accuracy, while groups with limited knowledge struggle to reach agreement, even after repeated discussion.}
    \label{fig:AccuracyWithTurns}
    \vspace{-1em}
\end{figure*}

\textbf{Contribution of Blockchain to Deliberation:}  
Deliberation is used prevalently in robotics \cite{swarmDeliberation}, governance \cite{GovernanceReview}, and Artificial Intelligence (AI) \cite{AgreeingOnPlansThroughIterated_Dispute}, highlighting its impact on improving decision-making performance. However, current deliberation implementations are limited in structured reasoning, argument verification, decision refinements, ensuring participation, and transparency \cite{morrell2005deliberation}. Blockchain's key feature of transparency and immutability takes deliberative consensus to another level. Not only do these features allow full visibility and analysis of how a certain decision was made, but they also allow the decision-making process to undergo public auditing. This is an excellent choice for reviewing decisions in critical sectors such as legal, healthcare, and government \cite{QuachSOSE}, \cite{Vinh}. 

\section{\textbf{Problem Definition}}
\vspace{-.5em}
The key idea of implementing a consensus mechanism in a decentralized blockchain network is to establish agreement and trust toward maintaining a unified and consistent ledger without a central authority \cite{Polkadot}. However, existing consensus algorithms are not designed to address the requirements of high-stakes applications that demand unanimous consensus with deliberation. For instance, a trial case in a courtroom requires all jurors to fully agree to decide whether someone is guilty before a verdict can be reached. With a limited agreement structure and no deliberation practice in current consensus protocols (e.g., PoW, PoS, PoS-BFT, DPoS, etc. \cite{pokharel2024deliberation}), it fails to allow enough coordination with participants. 
Furthermore, even though deliberation has been used to generate effective convergence in various real-life use-cases \cite{AgreeingOnPlansThroughIterated_Dispute}, \cite{DifferentLLMsUsedForDebate, acharya},  challenges including limited argument structure, lack of proposal refinements, biased or limited knowledge-based reasoning still hinder its ability to consistently produce fair and reliable outcomes for complex networks like blockchain \cite{Talapuru3D, TalapuruFostering}.

\section{\textbf{Contribution}}
\vspace{-.5em}
To the best of our knowledge, this is the first work to systematically identify the requirements of a deliberative unanimous consensus in blockchain networks for high-stakes applications, with the following contributions:

\begin{enumerate}
    \item Pioneering work using a multi-agent-based deliberation framework to achieve unanimous consensus in blockchain. \textbf{(See Section VI)} (\textbf{See Figure \ref{fig:RoundsOfDeliberation}})
    
    \item Formally defining the deliberative consensus protocol, and ensuring blockchain properties of consistency, agreement, and liveness hold under adversarial and stalled deliberation conditions.\textbf{(See Section XI-A)}

    \item Deliberation protocol evaluated on diverse datasets mirrors real-world deliberation where knowledgeable agents converge in fewer rounds, while less-informed ones struggle. \textbf{(See Section VIII, Figure 2)}
   
    \item The novel implementation work is open-sourced and made available \cite{blockchaincorejs}.
    
\end{enumerate}

\section{\textbf{Literature Review}}
\vspace{-.5em}
\subsubsection{\textbf{Cognitive and Social Models of Consensus}}
Negotiation is the process of bargaining with offers and counteroffers until an agreement is reached \cite{Negotiation}. In Argument-based negotiation, participants exchange "for" and "against" arguments to negotiate \cite{AgreeingOnPlansThroughIterated_Dispute, Paper1}. On the other hand, voting is a well-known method for reaching consensus \cite{Voting}. The work by Barabas makes the case that deliberation differs from discussion and other forms of consensus by allowing participants to soften strongly held views, encounter different perspectives, and learn readily \cite{Deliberation}. Deliberation's open-ended nature and expressiveness make it a better choice than negotiation and voting. 

\subsubsection{\textbf{Consensus on Blockchain}}
Lamport et al. addressed the byzantine generals problem to establish a distributed system where the number of honest nodes must satisfy n\textgreater 3t+1, where n is the total number of nodes and t is the number of dishonest nodes \cite{Laport}. Bitcoin later introduced a novel approach to financial transactions, ensuring finality through majority agreement among honest participants \cite{Bitcoin}. Even in fully honest networks, finality depends on majority agreement. Protocols like PoS and DPoS adopt this model by selecting validators for efficient finalization \cite{Polkadot}. This mechanism preserves system integrity while optimizing transaction processing speed to remain competitive with centralized systems. However, existing consensus models prioritize majority agreement over full participation, potentially overlooking critical perspectives. 
On the other hand, distributed applications such as Aragon Court \cite{DAOReview} and Kleros \cite{kleros_whitepaper} focus on fact-based adjudication rather than structured communication. Similarly, off-chain governance solutions like Decred’s Politeia \cite{Poletia} and Snapshot \cite{Snapshot} allow stakeholders to vote but lack structured deliberation. Moreover, reliance on off-chain discussions introduces transparency and security concerns, as critical decision-making steps occur outside the blockchain. 

\subsubsection{\textbf{LLMs and Deliberations}}
LLMs exhibit strong reasoning and generation abilities, enabling tasks like summarization, debate, and reflection with high accuracy \cite{LLMSurvey}. Techniques such as Chain of Thought \cite{ChainOfThought} decompose complex tasks into subtasks, inspiring collaborative approaches where multiple LLMs interact to enhance reasoning \cite{DifferentLLMsUsedForDebate}. These include cooperative \cite{improvingFractuality}, adversarial \cite{LLMROundTable}, or role-based settings \cite{HebarmasMachine}. However, these methods prioritize final outputs, discarding the deliberation process and thus lacking auditability and transparency. An in-depth overview of Multi-agent LLM systems is also presented in \cite{LLMSurvey}.  

\begin{figure*}[!t]
    \centering
    \begin{subfigure}[b]{0.29\linewidth}
        \centering
            \includegraphics[width=\linewidth]{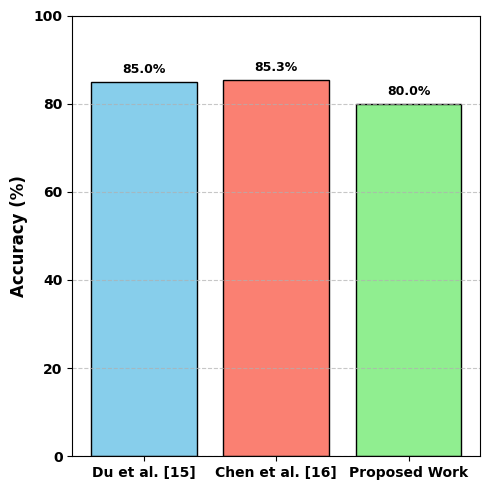}
        \caption{Comparison of Deliberation Protocol}
        \label{fig:llmComparision}
    \end{subfigure}
    \hfill
    \begin{subfigure}[b]{0.40\linewidth}
        \centering
        \includegraphics[width=\linewidth]{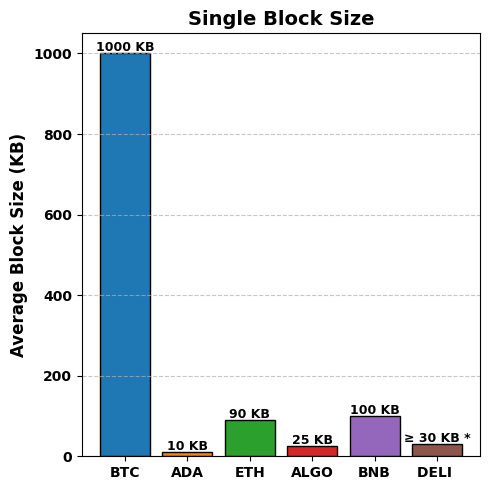}
        \caption{Block Size During Deliberation on Dataset \cite{cobbe2021gsm8k}}
        \label{fig:blocksize}
    \end{subfigure}
    \hfill
    \begin{subfigure}[b]{0.29\linewidth}
        \centering
        \includegraphics[width=\linewidth]{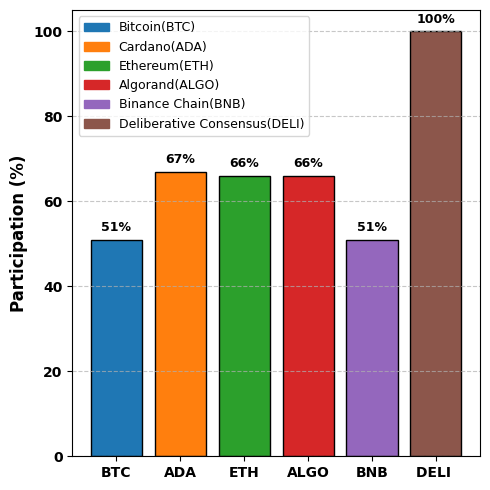}
        \caption{Node Participation \% for Consensus}
        \label{fig:blockchainComparision}
    \end{subfigure}
    \hfill
    \caption{a) \cite{improvingFractuality} uses an unspecified ChatGPT variant, \cite{LLMROundTable} uses GPT-4, Claude 2, and Bard. These models have larger parameter counts than the model employed in our work, which contributes to the observed differences in accuracy. b) Blocksize has a \textit{mild exponential} growth with number of agents, and \textit{linear growth} with number of rounds. 
    c) During unanimous consensus, all nodes participate and agree, addressing the issues with majority agreement that motivated this work.
}
    \label{fig:three-subfigs}
\end{figure*}

\section{\textbf{Prompt Engineering for LLM Optimization}}
\vspace{-.5em}
The utterance from an LLM greatly depends on the nature of the prompt given to it. To this end we use two kinds of prompting techniques defined below.

\subsubsection{\textbf{Initial Prompting}}
 The initial prompting phase (Algorithm \ref{alg:initial_prompting}) gathers each model’s (\(LLM_N\)) response \(r_N\) to the problem \(Pr\), storing them in the response set \(R^0\) for deliberation. To enhance realism and mimic real-world deliberation, we use a mix of Chain of Thought (CoT) prompts (\(P_{CoT}\)) and Zero Shot (ZS) prompts (\(P_{ZS}\)). CoT prompting improves response quality over ZS prompting \cite{ChainOfThought}.

\subsubsection{\textbf{Iterative Prompting/ Prompt Chaining}}
In order to allow change of opinions to take place the models are prompted in iterations (Algorithm \ref{alg:iterative_prompting}) where each iteration is defined as turns \(T\). In each turn \(t\), each agent receives its own utterance from the previous turn \(i\) as \(R_i^{t-1}\) and the previous turn's neighbor agent's responses as \(R^{t-1}\) as context and is tasked with evaluating and improving its own answer.

\begin{figure*}[htp!]
\centering
\includegraphics[width=.8\textwidth]{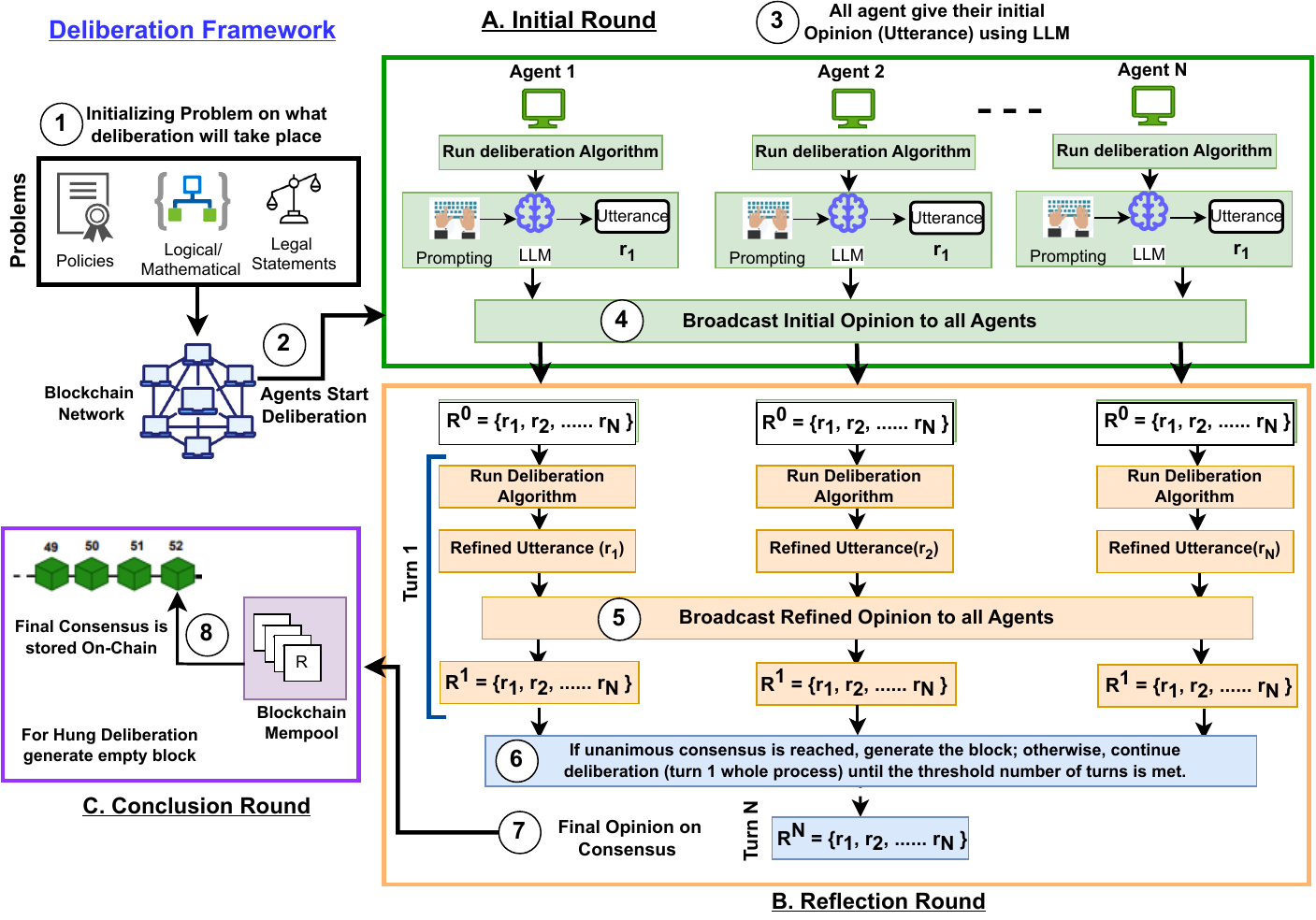}
 \vspace{-1em}
\caption{The figure shows the deliberation framework that starts with a problem. The \textit{initial round } gathers each agent's initial opinion on the problem. The \textit{reflection round} facilitates deliberation, allowing models to refine their viewpoints. Finally, the \textit{conclusion round} consolidates the refined utterances from the reflection round and publishes them on-chain.}  
\label{fig:RoundsOfDeliberation}
\end{figure*}

\section{\textbf{Deliberation Framework}}
\vspace{-.5em}
\label{DeliberationStructure}
To mimic real deliberation, the process unfolds over multiple rounds, as shown in Figure~\ref{fig:RoundsOfDeliberation}. We make an assumption that a solvable problem is selected for deliberation.

\subsubsection{\textbf{Initial Round}} 
For the initial round, initial prompting (Algorithm~\ref{alg:initial_prompting}) is used to collect individual perspectives from each model. Each utterance $(r_N)$ is stored in the response set $R^0$ and gossiped across the network. The number of agents ($N$) shapes the diversity of perspectives but increases storage overhead and prolongs convergence (see \ref{BlockSize}, Table \ref{tab:asd}).

\subsubsection{\textbf{Reflection Round}} 
For the second round, prompt chaining (Algorithm~\ref{alg:iterative_prompting}) allows agents to iteratively refine their responses over $T$ turns. Mimicking real-world deliberation, the process begins with diverse opinions and moves toward consensus. If unanimous agreement is reached, the round proceeds; otherwise, if convergence fails within $T$ turns, deliberation is declared hung and terminated. A round-robin strategy governs speaking order.

\subsubsection{\textbf{Conclusion Round}} 
For the final round of deliberation, a block is mined based on the outcome of the reflection round. If a unanimous consensus is reached, the full deliberation history is stored on-chain; otherwise, an empty block is mined to indicate a hung deliberation.

\section{\textbf{Integrating Deliberation in Blockchain}}
\vspace{-.5em}
This section describes how the underlying blockchain protocol is extended to support LLM-based multi-agent deliberation to generate a unanimous consensus.

\subsubsection{\textbf{Integrating LLMs in Blockchains}}
LLMs are a part of the blockchain node. Each node will be running its own LLM model and will call the model with prompts as mentioned in algorithm \ref{alg:initial_prompting} and \ref{alg:iterative_prompting}. Once the model replies, the utterance is propagated to every node in the network via a gossip protocol.

\subsubsection{\textbf{Consistency, Authenticity, and Propagation of Utterance}}
Utterances are signed by the producing agent and propagated via a two-phase gossip protocol. In phase one, nodes broadcast hashes of utterances; in phase two, peers request full data if needed. This reduces bandwidth while maintaining authenticity and consistency, as signatures are verifiable.

\subsubsection{\textbf{Storage, Verification and Extension of Ledger}}
Once consensus is reached, the complete deliberation is stored on-chain to ensure transparency and enable future dispute resolution. The resulting block is propagated across the network, allowing each model to verify its contents for integrity. Upon successful verification, the block is appended to the ledger and gossiped further. If consensus is not achieved within the predefined time $\mathcal{T}$, an empty block is mined, following the principle that "failure to agree is agreement to defer" \cite{Ripple}.

\section{\textbf{Experimental Results}}
\vspace{-.5em}
\label{Results}
\textbf{Implementation Detail:} This proof of concept extends the PoW client from Nimiq Blockchain\cite{NimiqWhitePaper} to build the proposed deliberation protocol. We use LLMs (Llama-3.1-70B-Instruct and Llama-2-13b-chat-hf) with NVIDIA RTX 6000 Ada Generation GPU. The GitHub repository is public \cite{blockchaincorejs}. The different dataset used are shown in table {\ref{tab:datasettable}}.

\subsection{\textbf{Impact of Deliberation}}
\vspace{-0.5em}
\label{DeliberationImpact}
\textbf{Time to Reach Consensus (TRC):}
Time taken to reach a unanimous consensus when varying agents, and turns are represented in Table \ref{tab:asd}. Table shows that the prompt generation time (PGT) is the \textit{primary contributor} to increasing TRC. Also, the number of agents and the number of turns directly correlate to more time due to more discussion taking place. 

\textbf{Relation Of Accuracy With Turns:}
 Figure \ref{fig:AccuracyWithTurns} shows consensus accuracy (correct agents / total) evolving through deliberation. Deliberating on the same problem, models with fewer parameters \textit{have low convergence and accuracy}, whereas larger model achieve \textit{higher accuracy and convergence}. 

\textbf{Comparison With Other Deliberation Protocols:}
 Figure \ref{fig:llmComparision} compares the accuracy of our protocol (80\%) with results reported in \cite{improvingFractuality} and \cite{LLMROundTable} on the GSM8K dataset. 

\subsection{\textbf{ Blockchain Performance}}
\vspace{-0.5em}
\textbf{Analysis of Block Size:}
\label{BlockSize}
Figure \ref{fig:blocksize} shows the blocksize obtained at the end of deliberation on the GSM8k dataset. The theoretical max size of our block is $2^{32}$ bytes. However, the size is limited to 100 kb. 

\textbf{Comparison with other Blockchain protocols:} A direct comparison is not feasible because block production time varies in our deliberative approach. During experiments, maximum block production took up to 150 seconds, with block sizes staying under 30KB. Current consensus approach require an honest majority; however, our deliberative protocol requires unanimity.  Figure \ref{fig:blockchainComparision} demonstrates this relationship.

\section{\textbf{Conclusion and Future Work}}
 \vspace{-0.5em}
This paper proposes a novel paradigm that integrates large language models (LLMs) within a blockchain to achieve unanimous consensus through deliberation. Our literature review highlights limitations in existing consensus protocols for decision-making and identifies gaps in the emerging field of LLM-driven blockchain consensus. To address this, we present a formal representation of using multi-agents to achieve unanimous consensus in decision-making through deliberation. Performance analysis on the proof of concept implementation work has also been presented. However, the system encounters potential challenges such as LLM reliability (hallucinations), security risks (bias, adversarial prompts), scalability (GPU and power demands), and Sybil attacks. These can be mitigated with validation, incentive mechanisms, efficient fine-tuning and off-chain storage, and reputation-based defenses respectively.
Future work will address the scalability of the deliberation framework, as well as explore prioritized deliberation more robustly.

\section{\textbf{Acknowledgment}}
\vspace{-0.5em}
This research was partially supported by NSA  grants H98230-20-1-0329, H98230-20-1-0414, H98230-21-1-0262, and H98230-22-1-0329. We thank Dr. Pascal Berrang from Nimiq, who helped us out initially with Nimiq's codebase.

\bibliographystyle{IEEEtran}
    \bibliography{ref}

@ARTICLE{ConsensusReviewPaper1,
  author={Lashkari, Bahareh and others},
  journal={IEEE Access}, 
  title={A Comprehensive Review of Blockchain Consensus Mechanisms}, 
  year={2021},
  volume={9},
  number={},
  pages={43620-43652},
  doi={10.1109/ACCESS.2021.3065880}}

@article{ConsensusReviewPaper2,
author = {Xu, Jie and others},
title = {A Survey of Blockchain Consensus Protocols},
year = {2023},
issue_date = {December 2023},
publisher = {Association for Computing Machinery},
address = {New York, NY, USA},
volume = {55},
number = {13s},
issn = {0360-0300},
url = {https://doi.org/10.1145/3579845},
doi = {10.1145/3579845},
journal = {ACM Comput. Surv.},
numpages = {35},
}

@article{cobbe2021gsm8k,
  title={Training Verifiers to Solve Math Word Problems},
  author={Cobbe, Karl and others},
  journal={arXiv preprint arXiv:2110.14168},
  year={2021}
}

@article{Negotiation,
	title = {Negotiation},
	
	
	number = {Volume 51, 2000},
	journal = {Annual Review of Psychology},
	author = {Bazerman, Max H. and others},
	year = {2000},
	
}

@article{Voting, 
    title={Why People Vote: Estimating the Social Returns to Voting}, 
    volume={46}, 
    journal={British Journal of Political Science}, 
    author={Gerber, Alan S. and others}, 
    year={2016}, 
   
}

@article{Deliberation,
	title = {How {} {Affects} {Policy} {Opinions}},
	volume = {98},
	issn = {0003-0554},
	url = {https://www.jstor.org/stable/4145332},
	number = {4},
	urldate = {2025-06-12},
	journal = {The American Political Science Review},
	author = {Barabas, Jason},
	year = {2004},
	pages = {687--701},
}

@INPROCEEDINGS{Paper1,
  author={Pokharel, Apurba and others},
  booktitle={6th International Conference on BCCA}, 
  title={ Leads to Unanimous Consensus}, 
  year={2024},
}

@article{morrell2005deliberation,
  title={Deliberation, democratic decision-making and internal political efficacy},
  author={Morrell, Michael E},
  journal={Political Behavior},
  volume={27},
  pages={49--69},
  year={2005},
  publisher={Springer}
}

@misc{DifferentLLMsUsedForDebate,
      title={Improving Language Model Negotiation with Self-Play and In-Context Learning from AI Feedback}, 
      author={Yao Fu and others},
      year={2023},
    
}

@inproceedings{ChainOfThought,
 author = {Wei, Jason and others},
 booktitle = {Advances in Neural Information Processing Systems},
 pages = {24824--24837},
 publisher = {Curran Associates, Inc.},
 title = {Chain-of-Thought Prompting Elicits Reasoning in Large Language Models},
 volume = {35},
 year = {2022}
}

@misc{improvingFractuality,
      title={Improving Factuality and Reasoning in Language Models through Multiagent Debate}, 
      author={Yilun Du and others},
      year={2023},
      url={https://arxiv.org/abs/2305.14325}, 
}

@misc{LLMROundTable,
      title={ReConcile: Round-Table Conference Improves Reasoning via Consensus among Diverse LLMs}, 
      author={Justin Chih-Yao Chen and others},
      year={2024},
      url={https://arxiv.org/abs/2309.13007}, 
}

@article{salerno2010promise,
  title={The promise of a cognitive perspective on jury deliberation},
  author={Salerno, Jessica M and others}, 
  year={2010},
  publisher={Springer}
}

@article{kleros_whitepaper,
  title={Kleros whitepaper},
  author={Lesaege, Cl´ement },
  journal={URL: https://kleros.io/static/whitepaper},
  year={2019}
}

@inproceedings{DAOReview,
author = {El Faqir, Youssef and others},
title = {An overview of decentralized autonomous organizations on the blockchain},
year = {2020}

}

@article{Poletia,
	title = {A novel framework for policy based on-chain governance of blockchain networks},
	volume = {58},
	journal = {Information Processing \& Management},
	author = {Dursun, Taner and others},
	year = {2021}
}

@online{Snapshot,
    author = {Snapshot Org},
    title = {Snapshot Docs},
    year = {2024},
    url = {https://docs.snapshot.org/introduction},
    note = {Accessed on April 9, 2024}
}

@online{blockchaincorejs,
    author = {Apurba Pokharel},
    title = {Blockchain-core-js},
    year = {2025},
    url = {https://github.com/apurbapokharel/blockchain-core-js},
}

@article{HebarmasMachine,
author = {Michael Henry Tessler  and others},
title = {AI can help humans find common ground in democratic deliberation},
journal = {Science},
volume = {386},
number = {6719},
pages = {eadq2852},
year = {2024},
}

@article{LLMSurvey,
	title = {A survey on {LLM}-based multi-agent systems: workflow, infrastructure, and challenges},
	shorttitle = {A survey on {LLM}-based multi-agent systems},
	url = {https://doi.org/10.1007/s44336-024-00009-2},
	journal = {Vicinagearth},
	author = {Li, Xinyi and others},
	pages = {9},
}

@INPROCEEDINGS{swarmDeliberation,
  author={Alhafnawi, Merihan and others},
  booktitle={2022 IEEE/RSJ IROS}, 
  title={Deliberative Democracy with Robot Swarms}, 
  year={2022},
  pages={7296-7303},
}

@inproceedings{zaman2024indivisible,
  title={Indivisible State Mirroring in Cross-Chain Atomic Swap across Heterogeneous Blockchains},
  author={Zaman, Shakila and others},
  booktitle={ 6th International Conference on BCCA}, 
  year={2024},
  organization={IEEE}
}

@inproceedings{zaman2023seamless,
  title={Seamless asset exchange in interconnected metaverses: Unraveling on-chain atomic swap},
  author={Zaman, Shakila and others},
  booktitle={5th IEEE TPS-ISA},
   year={2023},
 }

@inproceedings{pokharel2024deliberation,
  title={Deliberation leads to unanimous consensus},
  author={Pokharel, Apurba and others},
  booktitle={2024 6th International Conference on Blockchain Computing and Applications (BCCA)},
  pages={356--363},
  year={2024},
  organization={IEEE}
}

@article{GovernanceReview,
  author       = {Yue Liu and others},
  title        = {A Systematic Literature Review on Blockchain Governance},
  journal      = {CoRR},
  volume       = {abs/2105.05460},
  year         = {2021},
}

@article{AgreeingOnPlansThroughIterated_Dispute,
	title = {Agreeing on Plans Through Iterated Disputes},
	author = {Belesiotis, Alexandros and others},
}

@article{Laport,
author = {Lamport, Leslie and others},
title = {The Byzantine generals problem},
isbn = {9781450372701},
year={1982},
url = {https://doi.org/10.1145/3335772.3335936},
publisher = {Association for Computing Machinery},
}

@article{Ripple,
	title = {The {Ripple} {Protocol} {Consensus} {Algorithm}},
	language = {en},
	author = {Schwartz, David and others},	
}

@ARTICLE{NimiqWhitePaper,
	title = {Whitepaper},
	url = {https://www.nimiq.com/litepaper/},
}

@article{Bitcoin,
	title = {Bitcoin: {A} {Peer}-to-{Peer} {Electronic} {Cash} {System}},
	language = {en},
	author = {Nakamoto, Satoshi},
year={2008}
}

@misc{Algorand2,
      title={Algorand}, 
      author={Jing Chen and Silvio Micali},
      year={2017},
      url={https://arxiv.org/abs/1607.01341}, 
}

@article{Polkadot,
  title={Polkadot: Vision for a heterogeneous multi-chain framework},
  author={Wood, Gavin},
  journal={White paper},
  year={2016}
}

@online{GradeSchoolArithmetic,
    author = {Snapshot Org},
    title = {Snapshot Docs},
    year = {2024},
    url = {https://huggingface.co/datasets/georgiyozhegov/g.arithmetic},
}

@article{Delidata,
    title={DeliData: A dataset for deliberation in multi-party problem solving},
    author={Karadzhov, Georgi and others},
    journal={Proceedings of the ACM on Human-Computer Interaction},
    volume={7},
    pages={1--25},
    year={2023},
  }

@INPROCEEDINGS{acharya,
  author={Acharya, Dipak and Shu, Tong},
  booktitle={IEEE IPDPS}, 
  title={PredTOP: Latency Predictor Utilizing DAG Transformers for Distributed Deep Learning Training with Operator Parallelism}, 
  year={2025},
  pages={712-724}}

@INPROCEEDINGS{QuachSOSE, 
  author={Quach, Vinh and {et al.}}, 
  booktitle={2025 IEEE SOSE},  
  title={Access Microservices with Zero-Knowledge, Path Certainty and Distributed Traceability},  
  year={2025}
}

@INPROCEEDINGS{Vinh,
  author={Quach, Vinh and others},
  booktitle={ IEEE 6th TPS-ISA}, 
  title={ZCube: A Zero-Trust, Zero-Knowledge, and Zero-Memory Platform for Privacy and yet Secured Access}, 
  year={2024}
}

@INPROCEEDINGS{TalapuruFostering,
  author={Talapuru, Sirisha and others},
  booktitle={IEEE MetaCom}, 
  title={Fostering the Metaverse Immersion: Unraveling Personalized Dynamic Human Avatars}, 
  year={2024}
}

@inproceedings{Talapuru3D, 
author = {Talapuru, Sirisha and {et al.}}, 
title = {Bringing Life to 3D Human Avatars: Integrating Motion and Texture in Real-Time}, 
year = {2025}, 
booktitle = {Proceedings of the 30th International Conference on 3D Web Technology}, 
}

\section{Appendix}
\subsection{\textbf{Modeling Unanimous Consensus}}
\vspace{-.5em}
\textbf{Deliberation Game:}
    In order to prove the correctness of our deliberation framework, this section models our system as a cooperative game in which agents provide arguments that can be countered, refined, or enhanced. The end goal of this game is to arrive at a unanimous consensus through multiple iterations of the game rounds.
    
\begin{definition}
    \label{DeliberationGame}
    Let \(D\) be \(n\) player game where \(A_n\) represents the nth player, work together to solve a problem \(P\) by performing valid actions \(AC_r\)  in each iterative round \(r\) . The goal of the game is to achieve unanimous consensus with confidence \( C \) within a timebound \(\mathcal{T}\) with high confidence and the players receive a payoff \(PAY\) based on their contribution. \\
     Where, $D^i = (P, A_n, AC_r,PAY, C, t)$, 
    $D^i$ is the \(i^{th}\) deliberation done in the network, stored in a block $B$ and \(t\) is the time at which the deliberation completes.
    \end{definition}
    
\subsection{\textbf{Start of deliberation}}
\vspace{-.5em}

    \begin{definition}
    \label{StartOfDeliberation}
    The deliberation game \(D\) is said to be initiated if it meets one of the following criteria:
    \begin{enumerate}
        \item The problem \(P\) hasn't been deliberated before. 
        \item If deliberation was hung before $P \in H_g$. 
        Then $A_{n}^k != A_{n}^i$, where $i$ is the current deliberation, $A_{n}^k$ is the set of hung deliberators, and $, A_{n}^i$ is the new set of deliberators.
    \end{enumerate}
    \end{definition}
\subsection{\textbf{Unanimous Consensus Criteria}}
\vspace{-.5em}
    The type of consensus reached depends on the nature of the problem being deliberated on. Definitive problems have unique solutions, whereas problems involving prioritized preferences do not, giving rise to the notion of graded consensus \cite{Algorand2}.
    \vspace{-.5em}
    \begin{definition}
    \label{GradedConsensusForDefinitive}
    In a definitive deliberative game, \(D\) is in unanimous consensus if, for every n agents with m malicious agents, each honest agent \(h\) outputs a value-argument pair $AC_{r}^{i} = (V_r^i, Arg_r^i)$ at the end of round $r$ where the value is the answer to the problems along with arguments that support it, so that $\forall i, j \in h : \quad V_r^i = V_r^j$, \(C=1\) and \(t < \mathcal{T}\).
    \end{definition}

    \begin{definition}
     \vspace{-.5em}
    \label{GradedConsensusForDefinitive}
    In a prioritized deliberative game, we say that \(D\) is in unanimous consensus if, for every n agents with m malicious agents, each honest agent \(h\) outputs a policy $AC_{r}^{i} = \{Po_1^i, Po_2^i, ...Po_x^i\}$ at the end of round $r$ where \(Po_k\) are the policies proposed and are relevant to problem \(P\), so that $\theta$ is the fraction of honest agents that include policy \(Po_k\) in their action set and \(t < \mathcal{T}\).
    \end{definition}

\begin{definition}
\vspace{-.5em}
\label{AgreementConfidenceConsensus}
For a prioritized deliberation game, the confidence in the achieved consensus is measured as follows: Let $AC_{r}^{i} = \{Po_1^i, Po_2^i, \dots, Po_x^i\}$ be the set of policies proposed by agent $i$ at the end of round $r$.

For each unique policy \( Po_k \), its agreement level is defined as:
\begin{equation*}
A(Po_k) = \frac{\sum_{i \in h} \mathbb{1}(Po_k \in AC_r^i)}{|h|}
\end{equation*}
where \( \mathbb{1}(Po_k \in AC_r^i) \) is 1 if agent \( i \) included \( Po_k \) in \( AC_r^i \), and 0 otherwise. If the agreement level exceeds threshold \( \theta \), the policy is added to the accepted policy set:
\begin{equation*}
A(Po_k) \geq \theta, \quad Acc(Po) = Acc(Po) \cup A(Po_k)
\end{equation*}

The overall confidence in consensus is then computed as:
\begin{equation*}
T = \text{len}(Acc(Po)), \quad C = \frac{\sum A(Po_a)}{T} \quad \forall A(Po_a) \in Acc(Po)
\end{equation*}
where \( C \) represents the confidence in consensus.
\end{definition}

\subsection{\textbf{Termination of deliberation}}
\vspace{-.5em}
    \begin{definition}
    \label{termination1}
        The deliberation is said to be successfully terminated if the criteria for unanimous consensus are achieved. The deliberation parameters \(D^i\) are stored on the chain, and a new block is created.
    \end{definition}
     \vspace{-.5em}
    \begin{definition}
     \vspace{-.5em}
    \label{termination2}
        The deliberation is said to be in hung deliberation if: \(t > \mathcal{T}\) or if the number of honest participants falls below a threshold.
        The deliberation is then terminated by adding problem \(P\) to the hung problem set \(H_g\): \( H_g = H_g \cup P\) and producing an empty block along with the deliberation parameters \(D^i\).
    \end{definition}

\subsection{\textbf{System Properties}}
\vspace{-.5em}

\setcounter{subsubsection}{0}
\begin{theorem} Proof of consistency.
\subsubsection{\textbf{For definitive problem}}
For a definitive problem \(P\), initially honest player(s) \(l \in h\) will output a value-argument pair \((V_r, Arg_r)\) at the end of round \(r\) such that \((V_r)\) is same for all players \(l\) in the group.\\
After \(r_1, r_1>r\) rounds of deliberation all honest player \(i,j \in h\) and \(l < j,i\) outputs a value-argument pair \((V_{r_1}, Arg_{r_1})\) such that \( \forall i,j \in h\): \( V_{r_1}^i = V_{r_1}^j\). This implies that every honest player will output the same value ensuring consistency.
\subsubsection{\textbf{{For Prioritized Problems}}}
For a prioritized problem \( P \), initially, each honest agent \( i \in h \) outputs a set of proposed policies at the end of round \(r\):  $AC_r^i = \{ Po_1^i, Po_2^i, ..., Po_x^i \} $
where \( Po_k^i \) represents a policy proposed by agent \( i \) at the end of round \( r \). The confidence \(C_1\) in consensus is low.  

After \( r_1, r_1>r \) rounds of deliberation, every honest agent \( i, j \in h \) outputs a refined set of policies:  $ AC_{r_1}^i = \{ Po_1^{r_1}, Po_2^{r_1}, ..., Po_{T}^{r_1} \} $
such that the confidence \(C_2\) in consensus is higher than before \(C_2> C_1 \). This only occurs when the policies are selected consistently by more players at the end of the round \(r_1\).
Thus, for policies that are accepted and added to \( Acc(Po)\), the consistency property is maintained.
\end{theorem}
\setcounter{subsubsection}{1}
\begin{corollary} Proof of agreement.\\
The consistency property goes hand in hand with agreement. There can be no consistency without agreeing on the same values/policies, and no agreement without consistent outputs.
\end{corollary}

\begin{theorem} Proof of liveness \\
    Based on the definitions \ref{termination1} and \ref{termination2}, it can be stated that block production continues for both cases, ensuring liveness of the network.
\end{theorem}
 



\subsection{Miscelleaneous}
\vspace{-.75em}

\begin{table}[h]
    \centering
    \caption{Dataset used during Experimentation}
    \renewcommand{\arraystretch}{1.3} 
    \begin{tabular}{|p{1.5cm}|p{2.5cm}|p{2.5cm}|p{0.7cm}|}
    \hline
    \textbf{Dataset Name} & \textbf{Description} & \textbf{Size of Data} & \textbf{Refe-rence} \\
    \hline
    GSM8K &  Grade school math problems (algebraic) & 8,500 data & \cite{cobbe2021gsm8k} \\
    \hline
    DeliData  & Real-world deliberation on the Wason card problem & 500 data and each with up to 4 deliberators & \cite{Delidata} \\
    \hline
    G.arithmetic & Grade school math problems (arithmetic) & 1,000,000 data& \cite{GradeSchoolArithmetic} \\
    \hline
    \end{tabular}
    \label{tab:datasettable}
\end{table}

\vspace{-.75em}
\begin{table}[h]
    \centering
    \caption{Time for Unanimous Consensus (Dataset \cite{cobbe2021gsm8k})}
    \renewcommand{\arraystretch}{1.0}
    \begin{tabular}{|>{\centering\arraybackslash}p{.5cm}|
                    >{\centering\arraybackslash}p{.5cm}|
                    >{\centering\arraybackslash}p{.6cm}|
                    >{\centering\arraybackslash}p{.8cm}|
                    >{\centering\arraybackslash}p{1cm}|
                    >{\centering\arraybackslash}p{1.3cm}|
                    >{\centering\arraybackslash}p{.8cm}|}
        \hline
        \rowcolor{gray!30}
        {\textbf{\# Agent}} 
        & \textbf{\# Turns} 
        & \textbf{Model (billion)} 
        & \multicolumn{2}{c|}{\shortstack{\textbf{Latency (m:s)} \\ \textbf{With Prompt Generation}  }} 
        & \textbf{TRC (m:s)}
        & \textbf{PGT (m:s)} \\
        \hline
        & & & \textbf{IR} & \textbf{RR} & \textbf{IR + RR}& \\
        \hline

        3 & 2 & 70b & 1:33  & 1:00  & 2:33 & 1:55 \\
        3 & 3 & 70b & 1:31   & 1:20  & 2:51 &  2:05 \\
        4 & 2 & 70b & 2:08   & 1:00  & 3:08 & 2:09 \\
        4 & 3 & 70b & 2:02 & 2:15  & 4:17 & 3:12  \\
        5 & 2 & 70b & 2:32 & 2:10  & 4:42 & 3:32 \\
        5 & 3 & 70b & 2:37   & 2:55  & 5:32 & 3:51   \\

        \hline
    \end{tabular}
    \vspace{-0.5em}
    
    \label{tab:asd}
    {\footnotesize \textbf{Note: }IR: Initial Round; RR: Reflection Round.}
\end{table}

\vspace{-.75em}
\begin{algorithm}
\caption{Initial Prompting}
\label{alg:initial_prompting}
\SetAlgoNoLine
\SetKwInOut{Input}{Input}\SetKwInOut{Output}{Output}
\Input{Problem $Pr$, LLM, Num Agents $N$}
\Output{Initial Responses $R^0 = \{r_1, r_2, ..., r_N\}$}

$R^0 \gets \emptyset$;\\
\For{$i \gets 1$ \KwTo $N$}{
  $P_i \gets P_{CoT}$ if $i \in$ CoT group, else $P_{ZS}$;\\
  $r_i \gets LLM_i(P_i, Pr)$;\\
  Add $r_i$ to $R^0$;
}
\Return $R^0$
\end{algorithm}

\vspace{-.75em}
\begin{algorithm}
\caption{Iterative Prompting (Prompt Chaining)}
\label{alg:iterative_prompting}
\SetAlgoNoLine
\SetKwInOut{Input}{Input}\SetKwInOut{Output}{Output}
\Input{Question $Pr$, LLM, Num Agents $N$, Turns $T$}
\Output{Response sets per round $R^T$ and full set $R$}

Initialize $R^0$ using Algorithm~\ref{alg:initial_prompting};\\
\For{$t \gets 1$ \KwTo $T$}{
  \For{$i \gets 1$ \KwTo $N$}{
    $P_i^t \gets \text{Concat}(Pr, R_i^{t-1}, R^{t-1})$; \\
    $r_i^t \gets LLM_i(P_i^t)$; \\
    Add $r_i^t$ to $R^t$; where $R^t = \{r_1^{t}, r_2^{t}, ..., r_N^{t}\}$
  }
  Add $R^t$ to $R$; where $R = \{R^{0}, R^{1}, ..., R_N^{T}\}$
}
\Return $R$
\end{algorithm}


\end{document}